\documentclass[conference]{IEEEtran}
\bstctlcite{IEEEexample:BSTcontrol}
\ifCLASSINFOpdf

\else

\fi

\usepackage{graphicx}
\usepackage{subcaption}
\usepackage{float}
\usepackage{url}
\usepackage{amsmath,amssymb,epsfig}

\makeatletter
\let\old@ps@headings\ps@headings
\let\old@ps@IEEEtitlepagestyle\ps@IEEEtitlepagestyle
\def\confheader#1{%
	\def\ps@headings{%
		\old@ps@headings%
		\def\@oddhead{\strut\hfill#1\hfill\strut}%
		\def\@evenhead{\strut\hfill#1\hfill\strut}%
	}%
	\def\ps@IEEEtitlepagestyle{%
		\old@ps@IEEEtitlepagestyle%
		\def\@oddhead{\strut\hfill#1\hfill\strut}%
		\def\@evenhead{\strut\hfill#1\hfill\strut}%
	}%
	\ps@headings%
}
\makeatother

\confheader{%
	2019 13th International Conference on Mathematics, Actuarial Science, Computer Science and Statistics (MACS)
}

\usepackage[pscoord]{eso-pic}
\newcommand{\placetextbox}[3]{
	\setbox0=\hbox{#3}
	\AddToShipoutPictureFG*{ \put(\LenToUnit{#1\paperwidth},\LenToUnit{#2\paperheight}){\vtop{{\null}\makebox[0pt][c]{#3}}}
	}
}
\placetextbox{.50}{0.055}{\small{978-1-7281-4956-1/19/\$31.00 \copyright 2019 IEEE}}

\title{Heart Segmentation From MRI Scans Using Convolutional Neural Network}

\makeatletter
\newcommand{\linebreakand}{%
\end{@IEEEauthorhalign}
\hfill\mbox{}\par
\mbox{}\begin{@IEEEauthorhalign}
}
\makeatother
 		
    \author{
    	
    	\IEEEauthorblockN{Shakeel Muhammad Ibrahim}
    	\IEEEauthorblockA{\textit{Department of Computer Science,  } \\
    		\textit{University of Karachi,}\\
    		Karachi-75270, Pakistan. \\
    		shak.ibrhm.97@gmail.com}
    	\and
    	\IEEEauthorblockN{Muhammad Sohail Ibrahim}
    	\IEEEauthorblockA{\textit{College of Electrical Engineering} \\
    		\textit{Zhejiang University,}\\
    		Hangzhou, Zhejiang, China 310027. \\
    		msohail@zju.edu.cn}
    	
    	\and
    	\IEEEauthorblockN{Muhammad Usman}
    	\IEEEauthorblockA{\textit{FEST,} \\
    		\textit{Iqra University,}\\
    		Karachi-75500, Pakistan. \\
    		musman@iqra.edu.pk}
    	  \linebreakand
    	\and
    	\IEEEauthorblockN{Imran Naseem}
    	\IEEEauthorblockA{\textit{College of Engineering,} \\
    		\textit{Karachi Institute of Economics and Technology,}\\
    		Karachi 75190, Pakistan. \\
    		imrannaseem@pafkiet.edu.pk}
    	\and
    	\IEEEauthorblockN{Muhammad Moinuddin}
    	\IEEEauthorblockA{\textit{CEIES,} \\
    		\textit{King Abdulaziz University}\\
    		Saudi Arabia. \\
    		mmsansari@kau.edu.sa}  
}

\begin{document}   

\maketitle
\begin{abstract}
Heart is one of the vital organs of human body. A minor dysfunction of heart even for a short time interval can be fatal, therefore, efficient monitoring of its physiological state is essential for the patients with cardiovascular diseases. In the recent past, various computer assisted medical imaging systems have been proposed for the segmentation of the organ of interest. However, for the segmentation of heart using MRI, only few methods have been proposed each with its own merits and demerits. For further advancement in this area of research, we analyze  automated heart segmentation methods for magnetic resonance images. The analysis are based on deep learning methods that processes a full MR scan in a slice by slice fashion to predict desired mask for heart region. We design two encoder-decoder type fully convolutional neural network models (1) Multi-Channel input scheme (also known as 2.5D method), (2) a single channel input scheme with relatively large size network. Both models are evaluated on real MRI dataset and their performances are analysed for different test samples on standard measures such as Jaccard score, Youden's index and Dice score etc. Python implementation of our code is made publicly available at \url{https://github.com/Shak97/iceest2019} for performance evaluation. 

\end{abstract}

	
	\IEEEpeerreviewmaketitle

\section{Introduction}
\label{intro}
Heart is one of the vital organs of human body. It is the most hardworking organ that keeps circulating the blood in the body, thus playing indispensable role in maintaining the energy levels of the body. A minor dysfunction of heart even for a short time interval can be fatal, therefore efficient monitoring of its physiological state is essential for patients with cardiovascular diseases (CVD). To accomplish this important task, variety of medical imaging modalities have been introduced in the last few decades such as ultrasound, X-Ray computed tomography (CT), and magnetic resonance (MR) imaging etc.

The Ultrasound (US) is a real-time, radiation free and non-invasive and the safest imaging modality. However, it has some limitations such as, low contrast, perspective visualization (2D projection of 3D object), low penetration, and small field of view etc \cite{DNN_US}. On the other hand, CT provides 3D tomography with high quality speckle free images that have desired contrast information which can help in better diagnosis. But similar to  X-Ray, CT imaging is also a radiation based imaging therefore it is used as an offline imaging system that suits best in emergency situations. Due to its ionization characteristics, it is unsuitable for the patients who are at high risk e.g., pregnant women or young children etc \cite{robbins2008radiation}.

MR imaging (MRI) is the second most safest imaging modality. Unlike CT which works on the principle of radiation, in MRI, the body is magnetized by powerful magnets and the spin of the protons in hydrogen or other elements of choice (depending on the application of MRI) are synchronized with the resonant frequency. The change in the resonance of the spin is measured in Fourier domain and with the help of inverse Fourier transform high quality MR images are produced. To scan the region of interest (ROI), variety of scanning schemes are used which affect the quality, interpretation of pixel intensity, and acquisition time. By properly manipulating the control parameters, the desired quality of MRI can be obtained. A drawback of obtaining MRI is its scan time, which is several times higher than the CT or US imaging. However, for offline applications where highest contrast and superior resolution is required without the risk of radiation exposure, MRI is considered as a gold standard option. One such application is the detailed diagnosis of heart for its physical structural and metabolic functional analysis \cite{wang2019use}.  

With the advent of modern medical facilities it has now become an standard practice to store and analyse biomedical data to further improve the diagnostic processes. The type of data ranges from molecular level to the 3D MRI scan of fully human body.  Substantially, large medical imaging datasets are being introduced and the demand for automated medical assistance system is increasing subsequently. A thorough research in the domain of biomedical engineering has been observed in the recent years. The medical data is mostly heterogeneous as it is obtained after a variety of different clinical analysis procedures and the wide range of medical imaging modalities. Manual processing of such data is extremely difficult task which costs both the time and money to analyze the underlying information. In order to effectively utilize the acquired data, machine learning techniques have been effectively used for a wide number of biomedical applications such as protein prediction \cite{ECMSRC,RAFP_Pred,USAFP} and cancer classification \cite{khan2016novel,khan2017novel}.  

Various computer assisted medical imaging systems have been proposed for variety of biomedical applications such as classification of tumor, segmentation of organ of interest, and registration of multi-modality images etc. For computer assisted surgeries and diagnosis systems, the detection and identification of ROIs is one of many challenging tasks. The age, size and gender of the humans have direct impact on the  shapes and sizes of the internal organs. Therefore a great degree of variation is taken into account. The complications in identifying the upper and lower part of the heart can put even an experienced radiologist in a great inconvenience. 

Supplementary hitches are added while seeking computer based assistance.  To design a computer based medical imaging system, there is a need of high quality images in which ROIs can be easily distinguished e.g., in MRI. To the best of our knowledge only a few methods exist for the segmentation of heart \cite{xie2015holistically, baumgartner2017exploration, medical1}. Previously the segmentation task was performed using the conventional methods such as edge detection \cite{xie2015holistically}. 2D and 3D CNN were investigated for the tasks of heart segmentation in \cite{baumgartner2017exploration}. The use of few parameters were explored and it was revealed that the use of an optimized 2D approach yields better results than that of the 3D counterpart. More recently, in \cite{HS1} the authors proposed a shape reconstruction neural network (SCNN) and spatial constraint network (SCN) to achieve two different tasks. Firstly, SCNN was developed to maintain the shape of the heart in the segmentation results, secondly SCN was utilized to solve the problem of large variation in the 2D slices. In \cite{HS2} cardiopath classification of heart disease was performed. The ROIs were generated using the you only look once (YOLO) based object detection method. The proposed approach produced adequate accuracy results compared to the previous methods. A cluster based approach named vantage point is proposed in \cite {HS3}. Complete feature vector was used and representation pattern in higher dimensions feature space were found. 

For further advancement in this area of research, in this paper, we implemented an automated heart segmentation method for MR images. The proposed approach is a deep learning based method that processes a full MR scan in a slice by slice fashion to predict desired mask for heart region.

The contributions of this research are as follows:
\begin{enumerate}
	\item A fully automated computer assisted method is proposed for the segmentation of heart region from MRI.
	\item Two U-Net \cite{ronneberger2015u} like fully convolutional network models  are designed for the said task (i) multi-channel input scheme (also known as 2.5D method), (ii) a single channel input scheme with relatively large size network.
	\item A comparative analysis for both schemes are presented exhibiting their performances.
	\item  The proposed models are evaluated on real MRI dataset and their performance is analyzed for different standard measures such as Jaccard score and Dice score etc.
	\item Python implementation of our method is made public for the performance evaluation. 
\end{enumerate}

We organized this paper in following sections: in Section \ref{sec:method}, we discuss the details of our proposed method followed by the experimental results and discussion in section \ref{sec:results} and the conclusion of this study is presented in section \ref{sec:conclusion}.

\section{Method}
\label{sec:method}
The method is based on fully convolutional neural network, which is trained using the concept of data driven supervised learning, where representative examples are used to fine tune the learning parameter of the model. Typically medical imaging datasets are huge in size and have highly imbalanced distributions which make it difficult to design machine learning based model using conventional approaches. In this regard, the deep learning has proven to show great success \cite{medical2}.  MR scan produce 3D tomographic view of the body which is computationally very expensive to process with deep learning algorithms \cite{baumgartner2017exploration}. Therefore, to perform the heart segmentation from MRI scans, we divide 3D cubes into multiple 2D images.  For these 2D image planes, we trained a convolutional neural network that generates desired masks for each input image. Details of dataset, network configuration and training strategies are discussed in subsequent sections.

\subsection{Dataset}
\label{sec:dataset}
The dataset is downloaded from the medical segmentation decathlon (MSD) challenge dataset from (\url{https://decathlon-10.grand-challenge.org/home/}). The MSD challenge dataset is specially designed for generalized segmentation task and consists of 3D MR scans of $10$ different body parts including liver, pancreas and heart.  For this study, we used 'Task2' dataset which is comprised of $20$ heart MRI scans. Out of these $20$ scans, $14$ and $2$ scans are randomly selected for the training and validation of the model respectively, while remaining $4$ scans were used as test scans. Each 3D scan is converted into a set of images, having dimensions of $320\times320\times1$. In particular there are $1578$, $218$, and $455$ pairs of input/label images in training, validation, and test datasets respectively.

\subsection{Network Specification and Training}
\begin{figure}[h!]
	\begin{center}
		\centering
		\includegraphics[width=8cm]{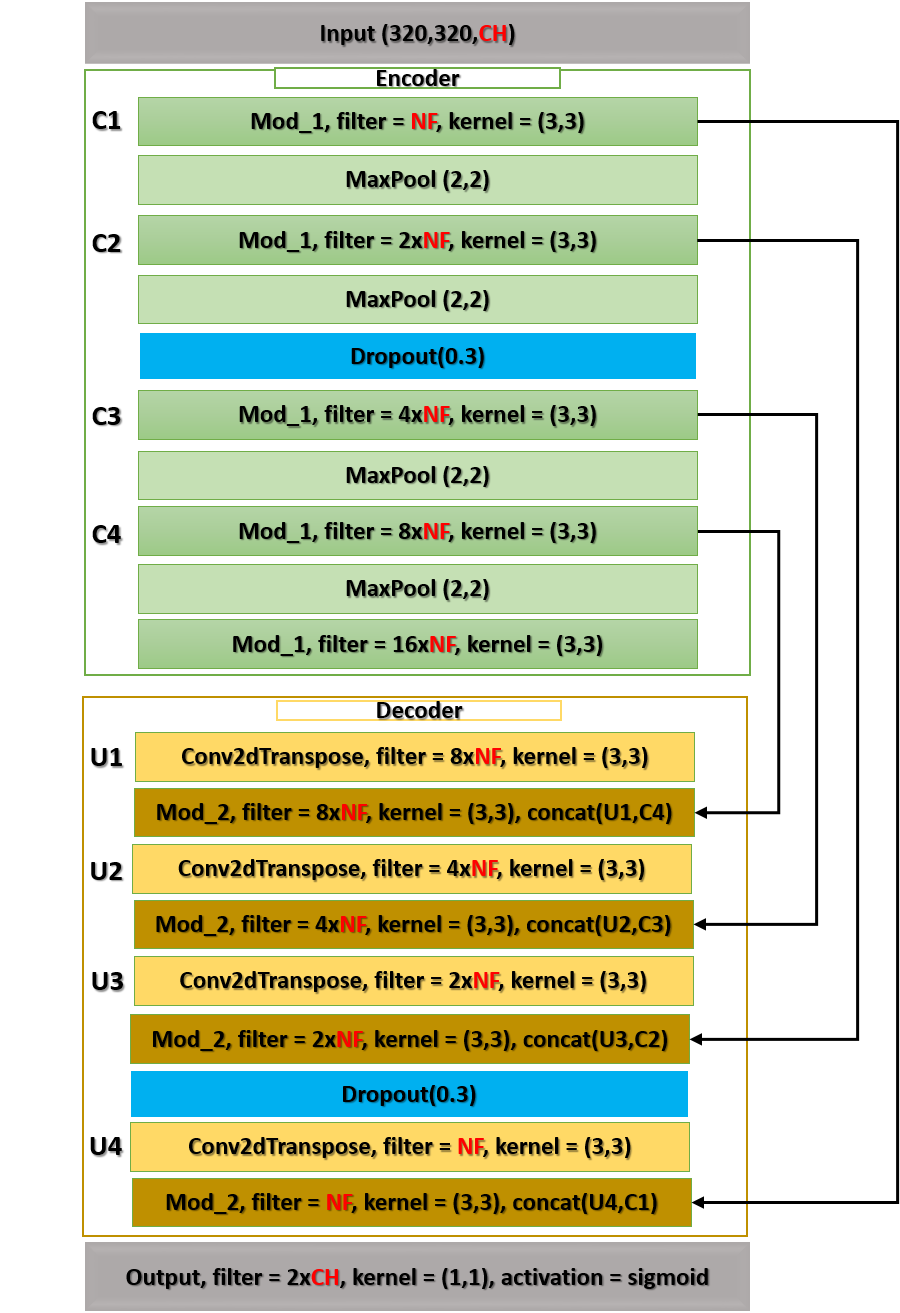}
	\end{center}
	\caption{Architecture of our convolutional neural network model.}
	\label{fig:cnn_architecture}
\end{figure}

\begin{figure}[h!]
	\begin{center}
		\centering
		\includegraphics[width=8cm]{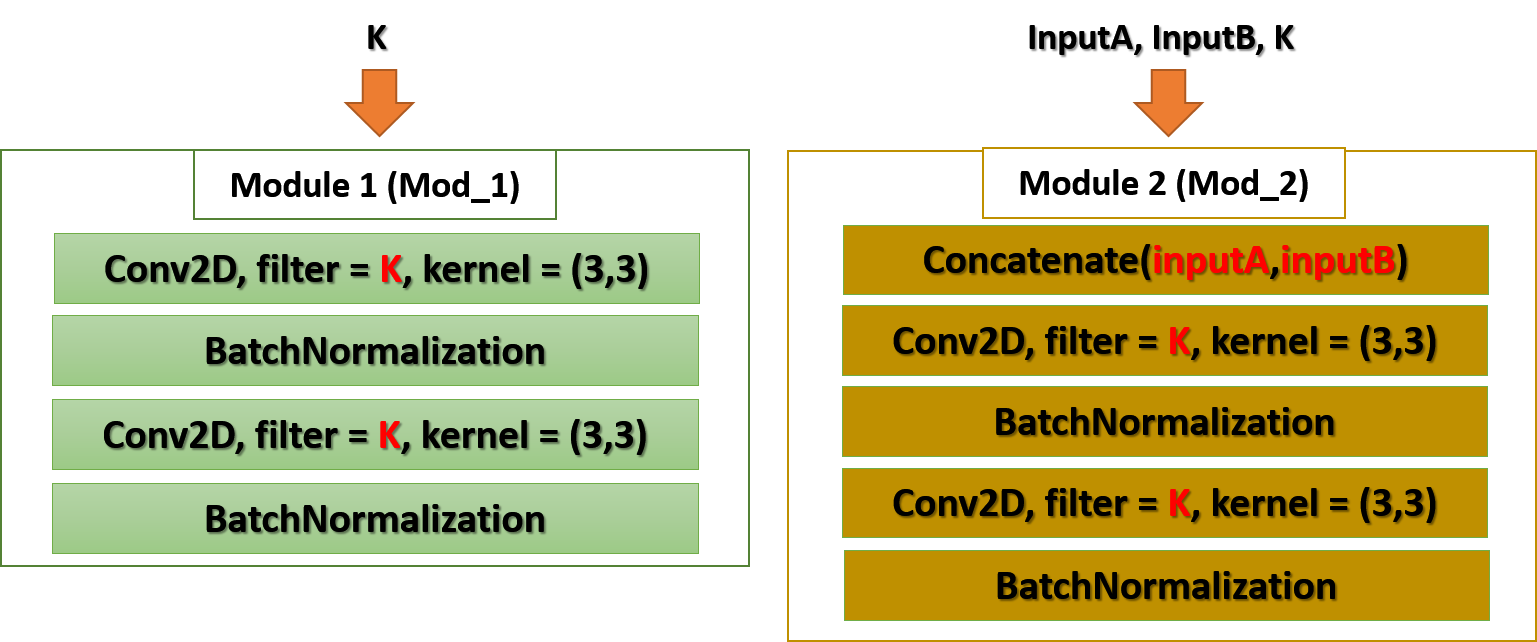}
	\end{center}
	\caption{Schematic diagram of individual modules of Fig. \ref{fig:cnn_architecture}}
	\label{fig:modules_block}
\end{figure}

We design $5$ different models which can be classified into two groups (1) thin networks with $1$, $3$, $5$, and $7$ input channels. (2) thick network with only $1$ input channel. All other hyper-parameters in both groups are same. Fig.\ref{fig:cnn_architecture} shows the generalized block diagram of models. The variable parameters are defined by number of input channels (CH) and number of filters (NF). As shown in Fig.\ref{fig:cnn_architecture}, the proposed design consists of two types of modules \textbf{Mod\_1} and \textbf{Mod\_2}, the configuration of which are depicted in  Fig.\ref{fig:modules_block}. In particular, \textbf{Mod\_1} consists of two convolution layers, two batch normalization layers and the $\textbf{K}$ number of filters, which for instance in $C3$ equals to $2\times$\textbf{NF}, whereas, \textbf{Mod\_2} contains an additional concatenation layer along with the two convolutional and batch normalization layers. As stated earlier the configuration of thin networks is $NF=16$ and $CH=\{1, 3, 5, 7\}$ and for thick network $NF=16$ and $CH=1$ is chosen. In all models,  ReLU activation function is used in every layer except the output layers where sigmoid function is used as an activation function. The Dropout and Max-pooling layers are also used as specified in Fig.\ref{fig:cnn_architecture}.

The model is implemented in Python TensorFlow using Keras library. To train a neural network, variety of algorithms have been proposed \cite{khan2018fractional, NNT1, NNT2}. Each has its own merits and demerits. However in general, the Adam optimizer \cite{Adam} is considered as the best choice for variety of problems. Therefore, to train our model we opt to use Adam optimizer, keeping the default value of all parameters. In the proposed method, we minimized the categorical cross entropy loss function $\mathcal{L}$ for $10$ epochs. The categorical cross entropy loss function is defined as
\begin{equation}
\label{equ:loss_function}
\mathcal{L}(y,\hat{y}) = -\sum_{r=0}^{R}\sum_{q=0}^{Q} (y_{q,r}\log(\hat{y}_{q,r})),
\end{equation}
 where $y$ is the original mask and $\hat{y}$ is the predicted mask in vector form, while Q is the number of training samples and R is the number of categories which is 2 in this study i.e., (\textbf{0}: background pixel and \textbf{1}: segment pixel). For each epoch, we monitor the training and validation losses along with the Dice scores (defined in \eqref{equ:F1}). 

\subsection{Performance Metrics}
We used following metrics to analyze the performance of the CNN models presented in this study.
\begin{enumerate}
    
    \item True positive rate (TPR), recall, or sensitivity
    \begin{equation}
\label{equ:TPR}
    TPR = \frac{TP}{TP + FN}
\end{equation}
    
    \item False positive rate (FPR), fallout, or (1- specificity)
    \begin{equation}
\label{equ:FPR}
    FPR = \frac{FP}{FP + TN}
\end{equation}

    \item Positive predicted value (PPV), or precision 
    \begin{equation}
\label{equ:PPV}
    PPV = \frac{TP}{TP + FP}
\end{equation}

    \item F1 / Dice score
    \begin{equation}
\label{equ:F1}
    Dice = \frac{2TP}{2TP + FP + FN}
\end{equation}

    \item Jaccard index (JI) or Jaccard similarity coefficient
    \begin{equation}
\label{equ:JI}
    JI = \frac{TP}{TP + FP + FN}
\end{equation}
 
    \item Youden index (YI) or informedness
    \begin{equation}
\label{equ:Youden}
    YI = TPR - FPR
\end{equation}
\end{enumerate}
where TP, FP, TN, and FP are the number of true positive, false positive, true negative, and false negative predictions of the number of pixels respectively.  

\section{Experimental Results and Discussion}\label{sec:results}
In this study we implemented five different models which are categorized into two groups (1) thin networks with $NF=16$ and $CH=\{1,3,5, 7\}$ and (2) thick network with $NF=64$ and $CH=1$. For all models, same datasets and hyper-parameters are used. The thin models with $CH=\{1,3,5, 7\}$ achieved dice score of $0.65$, $0.73$, $0.75$, and $0.77$ respectively. This indicates that the model with same number of learning parameters (NF), performs better if adequate neighbourhood information is provided. Although, the computational cost of the prediction is defined by the $NF$ and number of layers, however for each image we need to process $CH$ number of images simultaneously. Which means the memory requirement for each training epochs will increase. Another method to exploit the memory is to increase the size of network, which eventually increases the computational complexity.

\begin{figure}[h!]
	\begin{center}
		\includegraphics[width=8cm]{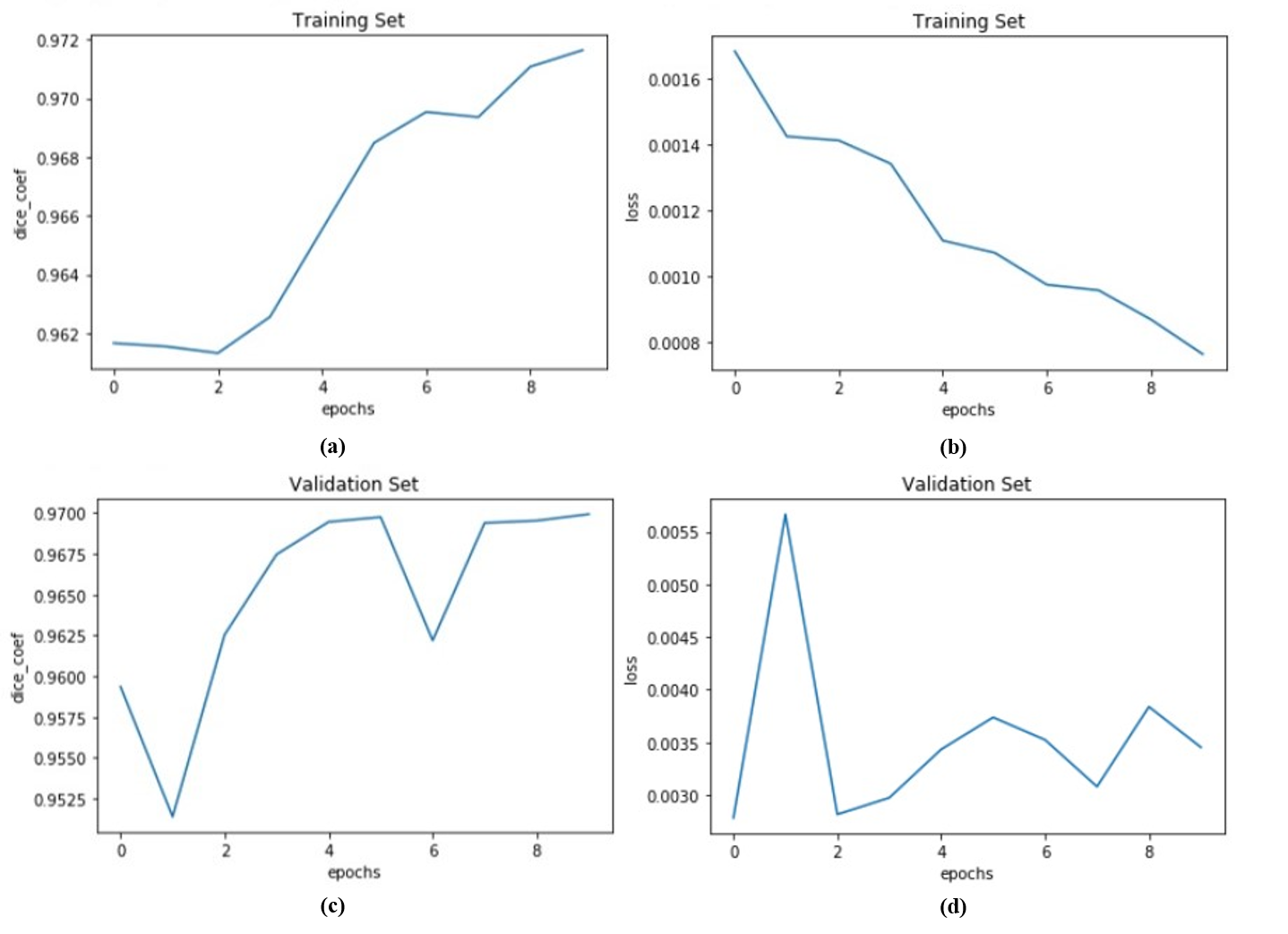}
	\end{center}
	\caption{(a) Dice coefficient curves for training dataset, (b) Cross entropy loss curve for training dataset, (c) Dice coefficient curves for validation dataset, (d) Cross entropy loss curve for validation dataset.}
	\label{fig:LDcurve}
\end{figure}

To analyze the effect of model's complexity, we design a relatively thick model with $NF=64$ and $CH=1$. This model presented significant performance improvement compared to the thin models. For better understanding of the performance we calculated variance statistics of our predicted masks and also analyze the visual difference in ground truth and prediction of the thick network.  Fig. \ref{fig:LDcurve} shows the dice coefficient and loss curves of 10 epochs for training and validation phases. Fig. \ref{fig:LDcurve}(a) shows the dice coefficient verses numbers of training epochs. A significant increase in the dice similarity is observed with subsequent increase in the number of epochs. The behavior of loss function with respect to the increase in epochs for the training phase is shown in Fig. \ref{fig:LDcurve}(b). The dice coefficient and loss function behavior of validation phase is depicted in Fig. \ref{fig:LDcurve}(c) and (d) respectively. Since no further increment in the validation performance is observed, therefore we stop the training after 10 epochs.

\begin{table}[!h]
	\begin{center}
		\caption{Performance statistics of model\#5}		\label{table:results}
		\begin{tabular}{|c|c|}
			\hline  \multicolumn{2}{|c|}{\textbf{Performance metrics}} \\
			\hline Recall & 0.7697 $\pm$ 0.3519 \\
			\hline Fallout & 0.0007 $\pm$ 0.0012 \\
			\hline Precision & 0.8471 $\pm$ 0.1602  \\
			\hline Dice score & 0.8216 $\pm$ 0.2067  \\
			\hline Jaccard index & 0.7361 $\pm$ 0.2260 \\
			\hline Youden's index & 0.7685 $\pm$ 0.3513 \\
			\hline 
		\end{tabular}
			\end{center}
\end{table} 

\begin{figure}[h!]
	\begin{center}
		\includegraphics[width=8cm]{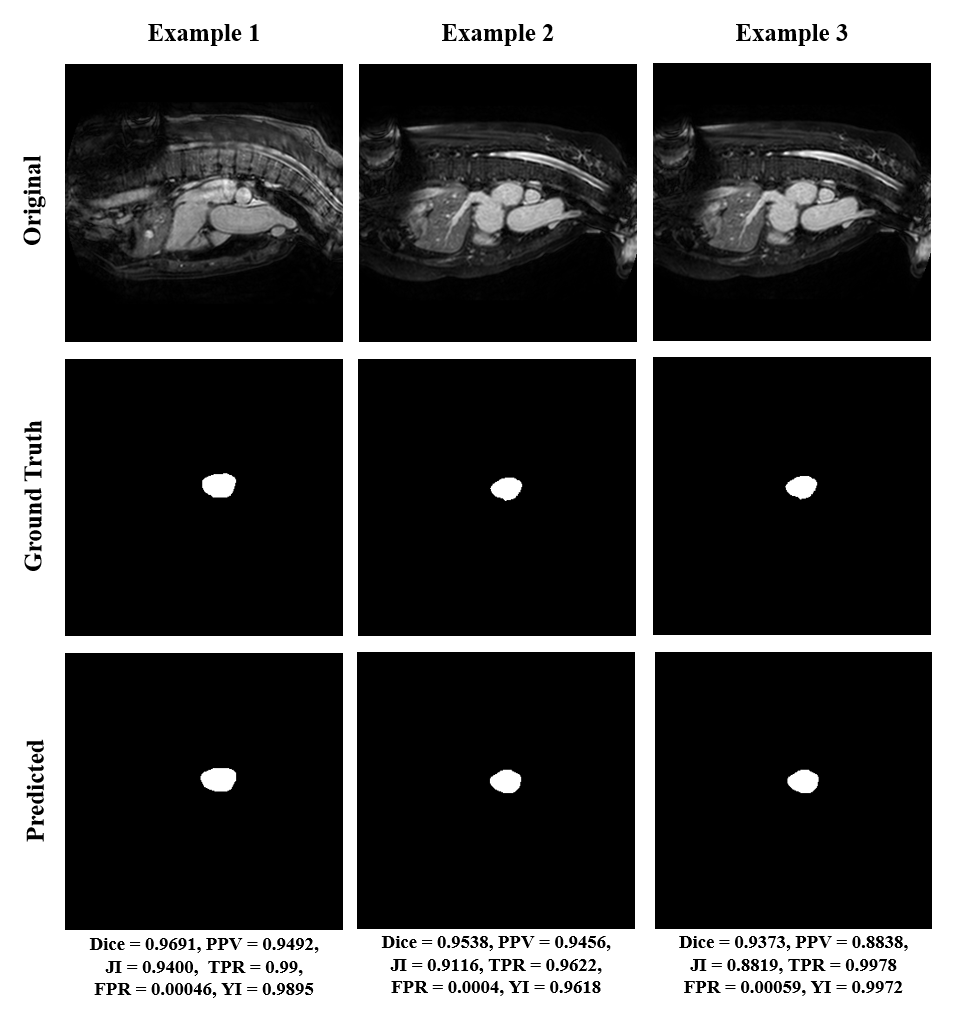}
	\end{center}
	\caption{Three good example results of heart segmentation using single channel CNN model}
	\label{fig: good}
\end{figure}

The visual examples of prediction are shown in Fig. \ref{fig: good}. It is evident from the figure that the ground truth has been accurately predicted, achieving a significant dice score of $0.973$, $0.953$, and $0.924$, for the examples 1, 2, and 3 respectively.

\begin{figure}[h]
	\begin{center}
		\includegraphics*[width=8cm]{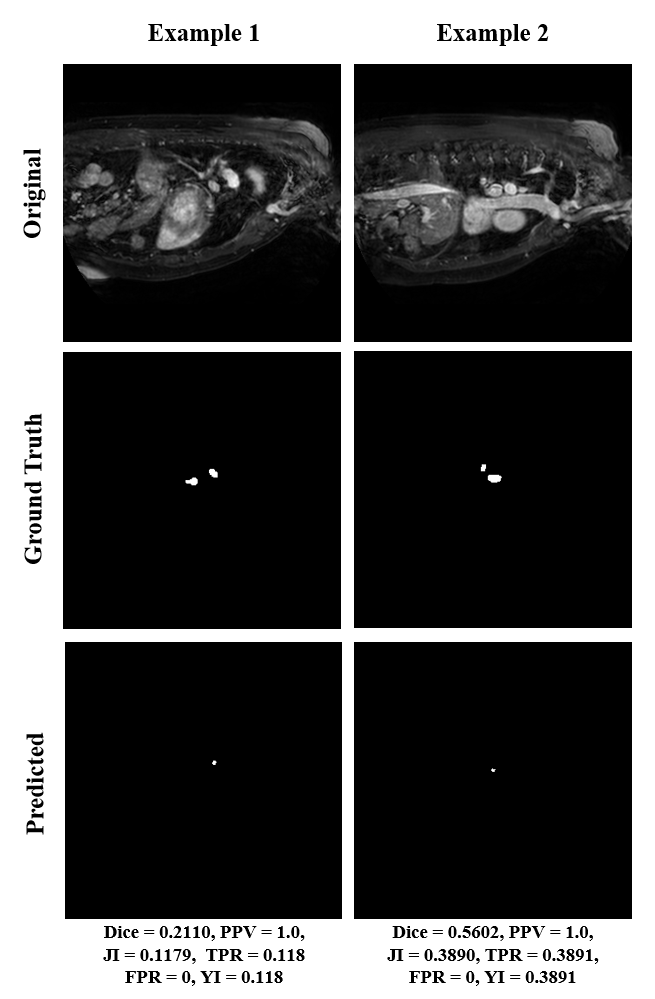}
	\end{center}
	\caption{Example results when single channel CNN model failed to predict correct mask.}
	\label{fig: bad}
\end{figure}

For the sake of comparison, some worst case scenarios in terms of prediction are also shown. These cases may occur due to various reasons such as over-fitting, inappropriate selection of input channels/filters, etc. Such schemes result in very low dice scores as shown in Fig. \ref{fig: bad}.  

This model yielded a dice score of $0.8216$ with a standard deviation of $\pm$ $0.2067$. The detailed statistical analysis is summarized in Table \ref{table:results}.

This clearly corroborates the proposition that a simple encoder-decoder styled CNN model with skip-connections can effectively predict the segmentation of heart region from MRI scans. It is noteworthy to point out that the the performance in large region's segmentation is very good however, for small sized segment the performance is very low which is also evident in inferior dice scores. 

\section{Conclusion}\label{sec:conclusion}
Herein, we demonstrated the application of deep learning for the segmentation of heart region from MRI scans. We design a fully convolution neural network based system which is trained using the concept of data driven supervised learning, where representative examples are used to fine tune the learning parameter of the model. Two approaches presented in this study revealed that a significant improvement in the performance can be achieved by either processing multiple images simultaneously through multiple input channels or by increasing the number of filters in each convolution layer. Increasing the input dimensions or the network parameters both contribute to the constructive performances. It can therefore be concluded that a simple encoder-decoder styled CNN model with skip-connections can effectively predict the segmentation of heart region from MRI scans, specially in the images with large segmentation region. However, for small segments the performance is not satisfactory, and therefore, the efficacy of the advanced CNN architectures and loss functions that can improve the accuracy for small size segments can be explored in future studies. 
	
	\bibliographystyle{IEEEtran}
	\bibliography{Reference}

\end{document}